\documentclass[twocolumn,nofootinbib,preprintnumbers,amsmath,amssymb]{revtex4}

\usepackage{graphicx}
\usepackage{dcolumn}
\usepackage{bm}
\usepackage{amsmath}

\def \be {\begin{equation}}
\def \ee {\end{equation}}

\begin{document}

\title{Pulsar Lightcurves}

\author{Andrei Gruzinov}

\affiliation{ CCPP, Physics Department, New York University, 4 Washington Place, New York, NY 10003
}

\begin{abstract}

Energy-resolved lightcurves are calculated for a weak pulsar. 

\end{abstract}

\maketitle

\section{Introduction}

Following the recipe of \cite{Gruzinov2013}, we calculate energy-resolved lightcurves for a weak pulsar\footnote{Here and below we allow ourselves to freely use the terminology of \cite{Gruzinov2013} because the present work makes no sense without \cite{Gruzinov2013} anyway: it would be odd to solve the more computationally demanding three-dimensional problem without first making sure that the axisymmetric calculation gives emission spectrum in approximate agreement with observations.}. We also confirm the recipe \cite{Gruzinov2013} by a full AE simulation. 

The calculation uses no adjustable parameters and, although at inadequate numerical resolution, yields the lightcurves similar to what is observed by the Fermi telescope \cite{Fermi2013}: compare Fig.(\ref{curves1}) and Fig.(\ref{curves2}) to the observed lightcurves of the pulsars listed in Table I of \cite{Gruzinov2013}. A better code (or just the same code with an increased resolution) should allow to measure both the observation angle and the spin-dipole angle for all observed weak pulsars by the lightcurve fitting.

In \S\ref{Lightcuves} we describe the calculation and results. In \S\ref{recipe} we confirm the recipe.

\section{Lightcurves}\label{Lightcuves}

We repeat the calculation of \cite{Gruzinov2013} in three dimensions, using a  three-dimensional version of the same (most primitive) code to solve Maxwell equations. The only differences are: (i) the rotating permanent (bulk) stellar current, (ii) the curvature of the charge trajectory, (iii) the calculation of the lightcurve.  We address these in turn.

(i) The external current is flowing toroidally around the axis $\hat{\bf{k}}$. The axis $\hat{\bf{k}}$ is inclined relative to the angular velocity of the star $\bf{\Omega }$, with $\hat{\bf{\Omega }}\cdot \hat{\bf{k}}=\cos \theta$, where $\theta$ is the spin-dipole angle. The current axis $\hat{\bf{k}}$ rotates with the angular velocity $\bf{\Omega }$. Another parameter of a pulsar is the observation angle $\chi$: $\cos \chi =\hat{\bf{\Omega }}\cdot \hat{\bf{n}}$, where $\hat{\bf{n}}$ is the direction to the observer.

(ii) The curvature, $K$,  of the trajectory of a charge moving with the unit velocity $\bf{v}$ is given by the acceleration: $K=|(\partial_t+{\bf v}\cdot \nabla){\bf v}|$. Since ${\bf v}={\bf v}({\bf E},{\bf B})$ is just an algebraic function of the local electromagnetic field, and the saturated electromagnetic field corotates with the star (while individual charges, of course, corotate only in the corotation zone), the velocity field ${\bf v}$ corotates too, i.e., $\partial _t{\bf v}=-{\bf V}\cdot \nabla {\bf v}+{\bf \Omega}\times {\bf v}$, where ${\bf V}={\bf \Omega} \times {\bf r}$ is the corotation velocity at the point with the radius vector ${\bf r}$. Thus the curvature is given by 
\be
K=|({\bf v}-{\bf V})\cdot \nabla {\bf v}+{\bf \Omega}\times {\bf v}|.
\ee

\begin{figure}[bth]
  \centering
  \includegraphics[width=0.48\textwidth]{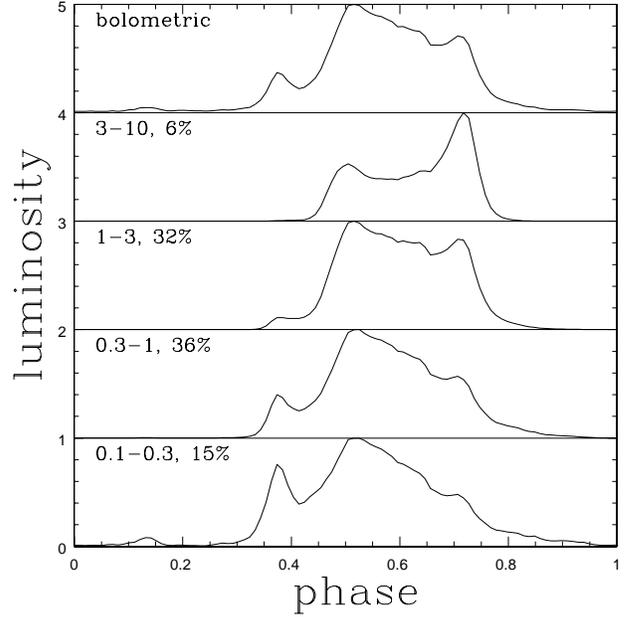}
\caption{Lightcurves of a weak pulsar; observation angle $\chi =110^{\circ }$; spin-dipole angle $\theta =45^{\circ }$. The labels indicate the photon energy intervals in arbitrary units and the fraction of power in these energy intervals.  } \label{curves1}
\end{figure}

(iii) The emission is very narrowly beamed along ${\bf v}$ and ${\bf v}$ corotates. It follows that one can calculate the observed time-dependent luminosity by integrating the radiation power (at fixed time) over $d^3r$ and sending each emitted photon into $\chi$, with $\cos \chi = \hat{\bf{\Omega }}\cdot {\bf v}$, and also assigning to each emitted photon a phase shift $\phi$. The phase $\phi$ consists of a time delay from the emission point to the observer, plus a time delay needed for the emitting point to rotate into a position at which it will illuminate the observer. To calculate $\phi$, put $\hat{{\bf \Omega }}=\hat{z}$, and place the observer into the $x$-plane. Then $\phi=-\tilde{\phi }-\Omega{\bf v}\cdot{\bf r}$, where $v_x=v_\perp \cos \tilde{\phi }$, $v_y=v_\perp \sin \tilde{\phi }$, $v_\perp=\sqrt{v_x^2+v_y^2}$.  One can then represent the entire family of lightcurves seen by different observers by $L(\phi ,\chi )$ -- the luminosity (bolometric or energy-resolved) observed at $\chi$ at a phase $\phi$ \cite{Bai2010}. In the plots, we follow the observers' convention and call ${\phi \over 2\pi}$ a phase.

With 270$^3$ grid, we place a star of radius $R_s=0.33$ at the center of a 6x6x6 box with outgoing boundary conditions. Based on what we have learned from the axisymmetric calculation \cite{Gruzinov2013}: the star is too big, the box is too small, the resolution is too low. Even though the results seem to make sense, we must state again that our accuracy should be poor. 

\begin{figure}[bth]
  \centering
  \includegraphics[width=0.48\textwidth]{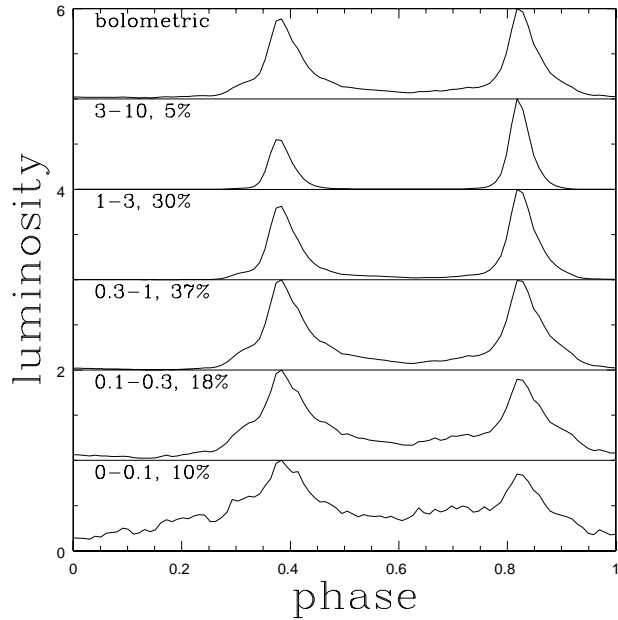}
\caption{Same as Fig.(\ref{curves1}); observation angle $\chi =96^{\circ }$.  } \label{curves2}
\end{figure}

\begin{figure}[bth]
  \centering
  \includegraphics[width=0.4\textwidth]{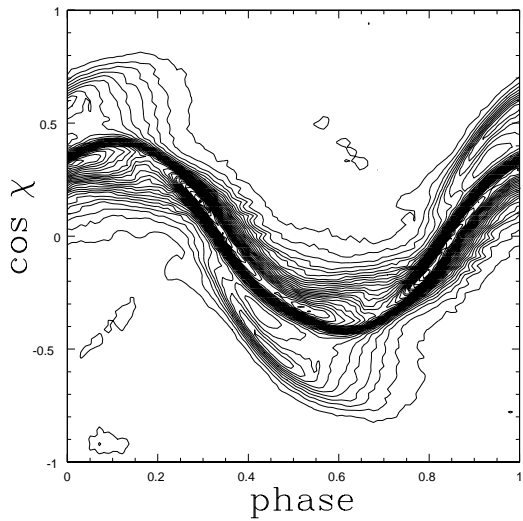}
\caption{Bolometric luminosity observed from $\chi$ at different phases.} \label{bol}
\end{figure}

\begin{figure}[bth]
  \centering
  \includegraphics[width=0.4\textwidth]{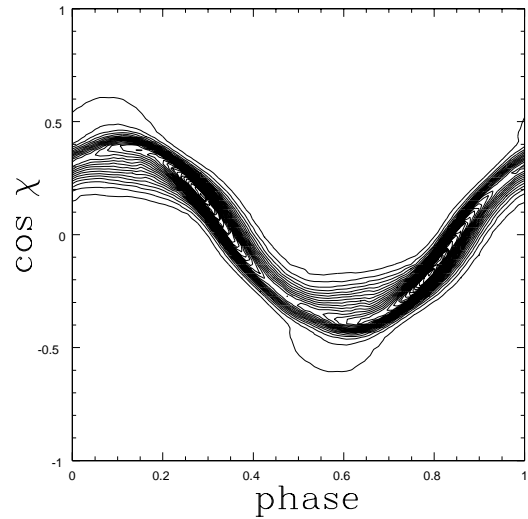}
\caption{Luminosity at photon energies greater than $E$, $L(>E)$, where $E$ is such that that the averaged, over observation angles and phases, $L(>E)$ is equal to $L_{\rm bol}/10$, where $L_{\rm bol}$ is the averaged, over observation angles and phases, bolometric luminosity.} \label{highe}
\end{figure}

\begin{figure}[bth]
  \centering
  \includegraphics[width=0.4\textwidth]{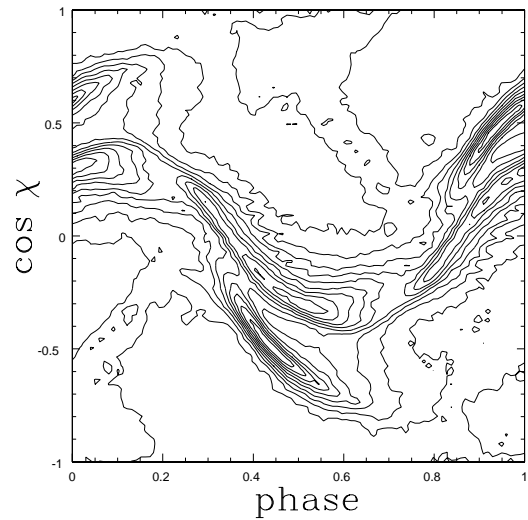}
\caption{Luminosity at photon energies less than $E$, $L(<E)$, where $E$ is such that that the averaged $L(<E)$ is equal to $L_{\rm bol}/3$.} \label{lowe}
\end{figure}

\begin{figure}[bth]
  \centering
  \includegraphics[width=0.48\textwidth]{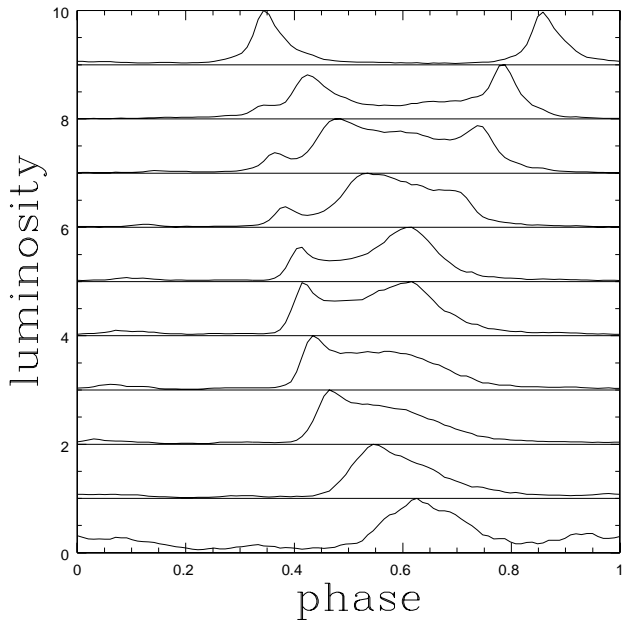}
\caption{Bolometric lightcurves; $\cos \chi$ decreases from 0 for the top curve to -0.76 for the bottom curve.} \label{bolcurve}
\end{figure}

Fig.(\ref{bol}) shows bolometric $L(\phi ,\chi )$ for a pulsar with the spin-dipole angle $\theta =45^{\circ }$. Fig.(\ref{highe}) shows the high-energy $L(\phi ,\chi )$ and Fig.(\ref{lowe}) shows the low-energy $L(\phi ,\chi )$ for the same pulsar. Fig.(\ref{bolcurve}) shows the bolometric lightcurves corresponding to Fig.(\ref{bol}). Figs.(\ref{curves1}, \ref{curves2}) show the energy-resolved lightcurves.

\section{Confirming the Recipe}\label{recipe}

\begin{figure}[bth]
  \centering
  \includegraphics[width=0.48\textwidth]{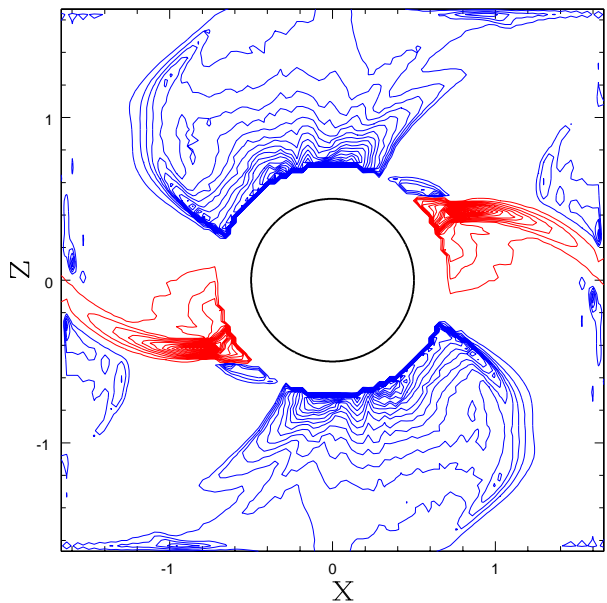}
\caption{Charge density (for $r>1.4R_s$); positive in red and negative in blue; computed by the recipe  \cite{Gruzinov2013}.} \label{den1}
\end{figure}

\begin{figure}[bth]
  \centering
  \includegraphics[width=0.48\textwidth]{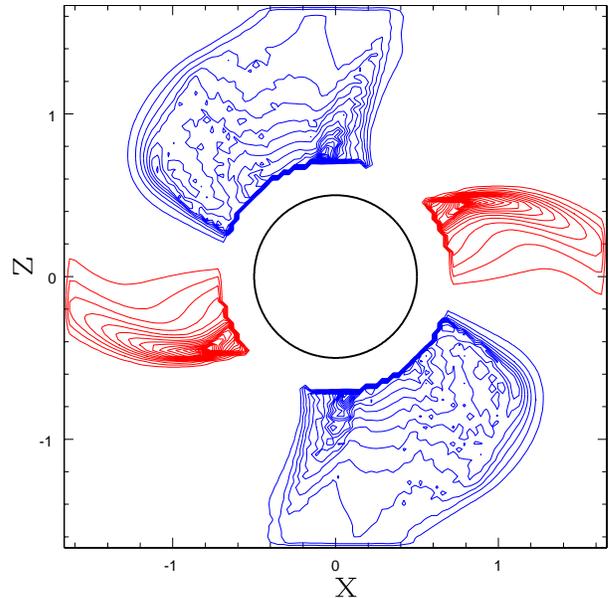}
\caption{Positron (red) and electron (blue) density (for $r>1.4R_s$); computed by full AE.} \label{den2}
\end{figure}

The recipe  \cite{Gruzinov2013} uses only the electromagnetic degrees of freedom plus some Ohm's law to calculate the magnetosphere. The particle densities are then deduced from the known electromagnetic field (at least everywhere in the radiation zone). Knowing the particle densities in the radiation zone, one calculates the emission.

That this procedure should give correct densities of charges for a weak pulsar is not at all obvious. The arguments of \cite{Gruzinov2013} are too elaborate to be fully trustworthy. Also, for non-weak pulsars, a full AE simulation, with explicit treatment of positrons and electrons seems to be a must. 

We therefore ran a three-dimensional full AE simulation, which describes electrons and positrons by their number density fields\footnote{ The code is just a three-dimensional version of the primitive code arXiv:1303.4094.}. The full AE simulation runs slower, so we limited ourselves by a 100$^3$ resolution, big star ($R_s=0.5$) and small box (the light cylinder of diameter 2 barely fits into the side 3.3 box). 

Now the resolution is totally inadequate. The radiation zone sits on the surface of our too big star. The influence of the box walls (which are obviously not perfectly outgoing) is too strong. Because of all these issues, we deliberately avoid any discussion of the most interesting part of the force-free zone, where, according to \cite{Gruzinov2013}, outgoing positrons should travel through a cloud of suspended electrons. (We do not even check our results there, plotting the charge density of the recipe run and the particle densities of the full AE run only down to 1/20 of the corresponding maximal values.)

However, we know (from calculating the radiation in this work and especially in \cite{Gruzinov2013}) that emission is heavily dominated by the region of enhanced positron density, shown in red in Figs.(\ref{den1}, \ref{den2}). And, at least in this region of high positron density, the full AE run does confirm the recipe \cite{Gruzinov2013} calculation. We then tentatively conclude that the recipe results are either exact or close to exact.

\section{Conclusion}
If we did not make a major error in numerics, and if better codes and higher resolution simulations do not discover new effects:

\begin{itemize}

\item AE fully solves the weak pulsar problem.
\item A simplified version of AE, the recipe \cite{Gruzinov2013}, is either exact or close to exact for weak pulsars.
\item One can measure the spin-dipole and observation angles of a weak pulsar.
\item To solve the non-weak pulsar problem, one needs to include pair production near the light cylinder. This might be doable. (Along the following lines: (i) use full AE to calculate electron and positron density, (ii) keep track of gamma-ray photons emitted by the curvature radiation, (iii) keep track of X-ray photons emitted by the synchrotron radiation of  newly born pairs and/or emitted by the star, (iv) add pairs and remove photons due to the X-ray-gamma-ray collisions.)

\end{itemize}

\end{document}